\documentclass[aps,prl,twocolumn,showpacs,superscriptaddress]{revtex4-1}

\usepackage{amsmath}
\usepackage{amssymb}
\usepackage{graphicx}
  \usepackage{hyperref}
\usepackage[arrow, matrix, curve]{xy}
\usepackage{bm}
\usepackage{yhmath}
\usepackage{color}
\newcommand\diff{\mathrm{d}}

\usepackage[usenames,dvipsnames]{xcolor}
\hypersetup{colorlinks=true, linkcolor=BrickRed, urlcolor=blue!50!black, citecolor=blue!50!black}

\usepackage[normalem]{ulem}

\makeatletter
\newcommand\hide@visible[1]{%
  \bgroup\fboxsep=.3ex\colorbox{Gray}{begin hide}%
  #1\colorbox{Gray}{end hide}\egroup%
}
\newcommand\hide@hidden[1]{%
  \bgroup\fboxsep=.3ex\colorbox{Gray}{hidden text}%
}
\newcommand\hide@invisible[1]{}
\newcommand\makevisible{\let\hide\hide@visible}
\newcommand\makehidden{\let\hide\hide@hidden}
\newcommand\makeinvisible{\let\hide\hide@invisible}
\makeatother
\makehidden

\begin{document}


\title{Fluids  in extreme confinement }



\author{Thomas Franosch}

\author{Simon Lang}
\affiliation{Institut f\"ur Theoretische Physik, Friedrich-Alexander-Universit\"at Erlangen-N\"urnberg, Staudtstra{\ss}e~7, 91058, Erlangen, Germany}


\author{Rolf Schilling}
\affiliation{Institut f\"ur Physik, Johannes Gutenberg-Universit\"at Mainz,
 Staudinger Weg 7, 55099 Mainz, Germany}



\date{\today}

\begin{abstract}
For extremely confined fluids with two-dimensional density $n$
 in  slit geometry of accessible width $L$, we prove that in the limit $L\to 0$
 the lateral and transversal degrees of freedom decouple, and the latter become ideal-gas-like. For small  wall separation the transverse degrees of freedom can be integrated out and renormalize
the interaction potential.
We identify $n L^2 $ as hidden smallness parameter of the confinement problem and evaluate the effective two-body potential analytically, which allows calculating
 the leading correction to
 the free energy exactly. Explicitly, we map  a fluid of hard spheres  in extreme confinement  onto a 2D fluid of disks with an effective
 hard-core diameter  and a soft boundary layer. Two-dimensional phase
transitions are robust and  the transition point experiences a shift   ${\cal O}(n L^2)$.
\end{abstract}






\maketitle


Confined fluids are intermediate between fluids in three and lower dimensions. The confinement strongly influences their physical behavior~\cite{Alba:2006} like structural~\cite{Dietrich:1995} and dynamical~\cite{Klafter:Restricted} properties, and in
particular the phase behavior and phase transitions~\cite{Evans:1990,Lowen:2009} or the glass transition~\cite{Lang:2010}. Therefore, confined fluids have attracted a lot of
attention during the last three decades. One of the widely discussed features is the influence of the restricted geometry on the critical behavior (see Ref.~\cite{Evans:1990,Binder:1974,Fisher:1981,Vink:2006,Liu:2010} and references therein). For a colloid-polymer mixture in a slit geometry with walls separated by five colloid diameters the critical exponents are already very close to those of the 2D liquid~\cite{Vink:2006}. Reduction of the spatial dimension from three to two replaces in the Kosterlitz-Thouless-Halperin-Nelson-Young (KTHNY) theory ~\cite{Kosterlitz:1973,Strandburg:1988} a first-order phase transition by a two-stage continuous transition from a fluid to a hexatic phase and then to a 'solid'  with long-range orientational, but algebraically decaying translational order \cite{Strandburg:1988,Dietrich:1995}.

Most studies consider a slit geometry with two parallel, hard plates separated by a distance $L+\sigma$ with a single component fluid of either  hard spheres with diameter $\sigma$ or point particles ($\sigma=0$).
We will consider the case of  $\textit{extremely}$ confined fluids where only a monolayer fits between the walls. Most investigations addressing this regime used computer simulations (see Refs.~\cite{Schmidt:1996,Schmidt:1997,Gribova:2011} and references therein), density functional theory (DFT)(see Refs.~\cite{Tarazona:1987, Rosenfeld:1996, Rosenfeld:1997, Goetzelmann:1997}), virial expansion, free volume theory, effective diameter theory~\cite{Schmidt:1997}, integral equations~\cite{Henderson:1986,Adams:1989,Xu:2008a,Xu:2008b}, and experiments~\cite{Nygard:2009}. For instance, the density profile at the center
of a neutral hard-sphere fluid between two parallel neutral hard walls (HSHW) has been calculated exactly for $L/\sigma\to 0$~\cite{Henderson:1986, Goetzelmann:1997}. How the 3D functional for the excess free energy reduces to the corresponding 2D functional was investigated for a HSHW for $0\leq L \leq \sigma $ in Refs.~\cite{Tarazona:1987, Rosenfeld:1996}. 
The approximate analytical approach and the computer simulations~\cite{Schmidt:1996,Schmidt:1997} as well as experiments~\cite{Pieranski:1983,Neser:1997}  of a HSHW reveal that the transition of the 2D system from the fluid to the 'solid' triangular phase and from the latter to a buckling phase, a stable bilayer phase first observed and theoretically explained in Refs.~\cite{Bonnisent:1984,Pieranski:1983},  exists up to $L/\sigma \simeq 0.6$. For $L/\sigma \gtrsim 0.6$ new phases appear which do not have an analogue in the '2D world'. That the hexatic phase (probably not found in Refs.~\cite{Schmidt:1996,Schmidt:1997} because the system is too small) exists for $L/\sigma_{\text{LJ}} \leq 0.15$ has been shown  recently  for a Lennard-Jones (LJ) liquid of particles with a 'diameter' $\sigma_{\text{LJ}}$~\cite{Gribova:2011}.

Intuitively it is obvious that a fluid confined between two plates approaches a   2D fluid as $L\to 0$. But  one of the interesting questions is: How does such a fluid converge to a 2D fluid if the effective distance $L$ becomes smaller and smaller? Or vice versa: If a 2D fluid, e.g. undergoes an equilibrium phase transition, what is the range of $L$ such that the transversal degrees of freedom (d.o.f.)  do not affect the properties of that  transition? To study these questions is  the main motivation of the present work. Surprisingly, analytically exact results can be derived, which is a rare situation for strongly interacting many-particle systems.
We will show that for $L\to 0$ the lateral d.o.f. decouple from the transversal ones, where the latter behave as an ideal gas in the external wall potential. Additionally we will calculate the exact leading order correction due to their coupling.  This allows us to determine the leading $L$ dependence of thermodynamical quantities.

We investigate a fluid of $N$ identical particles with lateral and transversal d.o.f.  $\vec{x}_i = (\vec{r}_i, z_i)$, $i=1,2,\ldots,N$. The fluid is
confined between two plates at $z=\pm (L+\sigma)/2$ parallel to the x-y plane, and
the area of a plate is denoted by $A$.
 The particles mutually  interact via a pair potential only
\begin{align}\label{eq:potential}
V(\{ \vec{x}_i \}) &= \sum_{i<j} {\cal V}(x_{ij}),
\end{align}
where we abbreviate $x_{ij} := |\vec{x}_i-\vec{x}_j|$. To illustrate our approach we consider a pure hard-core repulsion with
  core diameter $\sigma$ and ignore additional particle-wall interactions. A generalization to smooth pair and particle-wall interactions is straightforward. As usual the configurational partition function reads $Z= {\sf Tr}\left( \exp\left[-\beta V( \{ \vec{x}_i \})\right]\right) $ with
the configurational integral  ${\sf Tr}( \cdot ) = \int \left[ \prod_{i=1}^N  \diff^3 x_i \right] (\cdot )$.  Then $F_{\text{ex}} =  - k_B T \ln (Z/V^N)$ is the excess free energy~\cite{Hansen:Theory_of_Simple_Liquids}
 with respect to a three-dimensional ideal gas of accessible volume $V= A L$.

Let us outline the strategy of our approach. First, it will be shown that the configurational part $\rho( \{ \vec{x}_i \})=Z^{-1} \exp\left[-\beta V( \{ \vec{x}_i \})\right]$ of the canonical ensemble
factorizes for $L \to 0$ into a transversal and lateral distribution. In a second step a cluster expansion with respect to a 2D reference fluid is developed, which allows us to eliminate the transversal d.o.f. and to obtain an effective potential $V_\text{eff}(\{\vec{r}_i\};L)$, that adds to the bare potential of the reference fluid. Third,
we employ
 $V_\text{eff}(\{ \vec{r}_i\};L)$ to calculate the leading correction to the free energy.

\begin{figure}
\includegraphics[angle=0,width=0.9\linewidth]{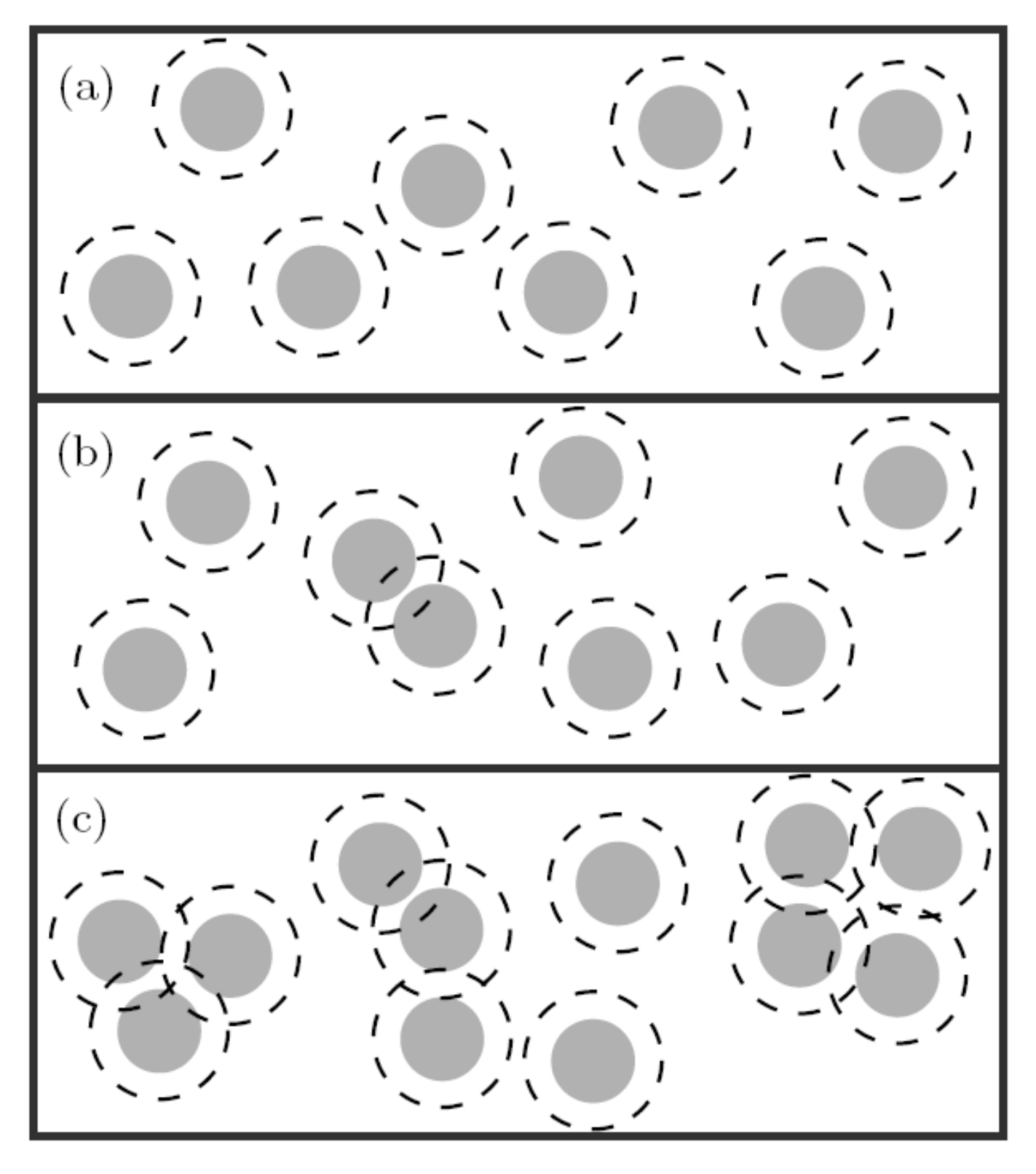}
\caption{Illustration of the cluster expansion: The hard core with diameter $\sigma_{L}$ is shown in gray and the dashed circle with diameter
$\sigma$ marks the range above which the excluded volume interactions vanish. 2D fluid of hard disks with diameter $\sigma_{L}$ (a),
a $2$-cluster (b), two irreducible $3$-clusters and an irreducible $4$-cluster (c) immersed in the hard disk fluid.}
\label{fig:cluster_expansion}
\end{figure}

The key observation is that by
 pure geometrical reasons
 the lateral coordinates of  two neighboring spheres cannot come closer
 than  $\sigma_{L}= \sqrt{ \sigma^2- L^2}$  for small $L$ as follows directly by Pythagoras' theorem.
Hence the effective interaction after tracing out the transversal degrees of freedom is concentrated on a thin shell of area $\pi (\sigma^2 -\sigma_{L}^2)=\pi L^2$ in addition to the bare two-dimensional hard-core repulsion of diameter $\sigma_L$, to be denoted by $W(\{\vec{r}_i\};L) = \sum_{i< j} {\cal W}(r_{ij};L)$. To avoid cumbersome notation, we mostly suppress the explicit dependence on $L$ in the following for the interaction potentials.  Then the distribution function (as a measure) factorizes to leading order
$\rho( \{ \vec{x}_i \} ) = \rho_\perp(\{ z_i \} ) \rho_\parallel(\{ \vec{r}_i \}) [1 + {\cal O}(L^2)]$
into the distribution function of the lateral and transversal d.o.f..  Here the transversal distribution reduces to a one-dimensional ideal gas, $\rho_\perp(\{ z_i \} ) = 1/Z_\perp$, and trivial partition function $Z_\perp = L^N$, whereas the lateral d.o.f. correspond to a two-dimensional hard-disk fluid $\rho_\parallel(\{ \vec{r}_i \})
= \exp(-\beta W( \{ \vec{r}_i \}))/Z_\parallel$. In particular, the free energy simplifies to $F= F_{\text{id}} + F^\parallel_\text{ex} + {\cal O}(L^2)$  where  $F_{\text{id}}$ is the three-dimension ideal gas contribution due to the kinetic energy,  and $F^{\parallel}_\text{ex} = - k_B T \ln (Z_\parallel/ A^N)$ is the excess free energy of a hard-disk fluid of diameter $\sigma_L$.

\begin{figure}
\includegraphics[angle=0,width=0.9\linewidth]{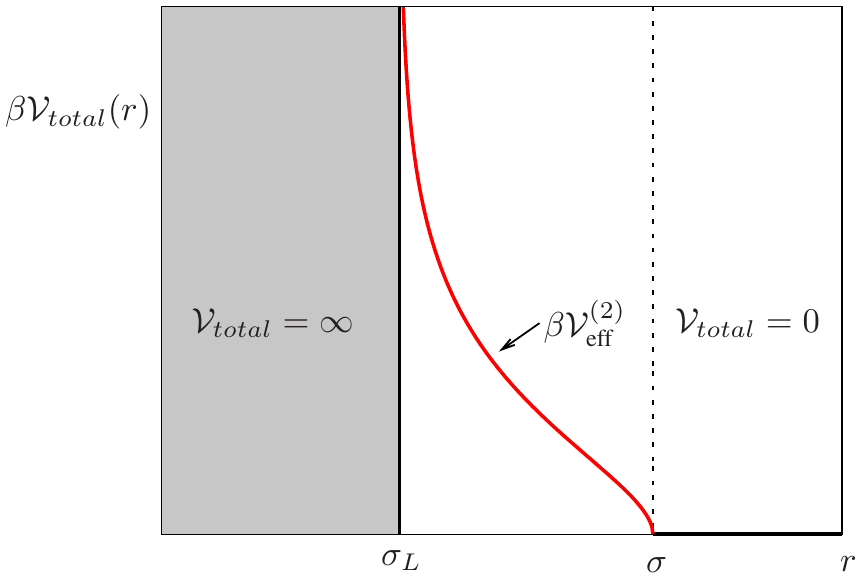}
\caption{The total two-body potential  as function of distance $r$. The gray region represents the excluded volume region corresponding to the closest lateral  distance $\sigma_L$ two spheres can assume in confinement. }
\label{fig:soft_potential}
\end{figure}

Next, we elaborate the leading correction to the factorized ensemble. Let us introduce the cluster functions $f_{ij}\equiv f(r_{ij},z_{i},z_{j}) = \Theta(r_{ij}^2+(z_{i}-z_{j})^2-\sigma^2)-\Theta(r_{ij}^2-\sigma_{L}^2) $ with $\Theta(x)$ the Heaviside function,
and $r_{ij} := |\vec{r}_i - \vec{r}_j|$. Note, $-1 \leq f_{ij} \leq 0$ and its support is restricted to $\sigma_L < r_{ij} < \sigma$.  Then the identity  $\exp[-\beta V(\{ \vec{x}_i\})] =\exp[-\beta W(\{ \vec{r}_i\})]\prod_{i<j}(1+f_{ij})$ allows us to perform a cluster expansion. Let us emphasize that the subsequent procedure can be directly generalized to the case of additional smooth pair and particle-wall interactions or point particles by suitable choice of  the reference potential $W(\{ \vec{r}_i\})$ and cluster functions $f_{ij}$. For convenience we abbreviate  pairs $i\neq j$ by $\alpha = (ij)$ and enumerate them. Then, we define the effective potential $\exp(-\beta V_\text{eff} )  = \langle \prod_\alpha (1+ f_\alpha ) \rangle_\perp$ by averaging over the transversal d.o.f. . With $x= \langle \prod_\alpha (1+ f_\alpha ) \rangle_\perp -1$ and using the series expansion $\ln (1+ x) = \sum_{k=1}^\infty (-1)^{k+1} x^k/k$, one finds
\begin{align}\label{eq:f}
 -\beta V_\text{eff}  
=&  \sum_\alpha \left[\langle f_\alpha \rangle_\perp - \frac{1}{2} \left( \langle f_\alpha \rangle_\perp \right)^2 +\ldots \right]\nonumber \\
 &+ \sum_{\alpha < \beta } \left[\langle f_\alpha f_\beta \rangle_\perp - \langle f_\alpha \rangle_\perp \langle f_\beta \rangle_\perp \right] +\ldots
 \end{align}
The first line contains precisely the terms $(-1)^{k+1} \sum_\alpha [\langle f_\alpha \rangle_\perp ]^k/k$ and adds up to the exact effective two-body potential $\mathcal{V}_{\text{eff}}^{(2)}$, the subsequent term contains the first contribution to the three-body interaction,
\begin{equation}\label{eq:effective}
  V_\text{eff} = - k_B T \sum_\alpha \ln (1+\langle f_\alpha \rangle_\perp ) + \sum_{k=3}^N {\cal O}(\text{k-clusters}).
\end{equation}
Note that  successive cluster contributions are additive and involve $k$-body interactions $\mathcal{V}_{\text{eff}}^{(k)}$, see  Fig. \ref{fig:cluster_expansion} for illustration of the various clusters.
Keeping only the two-cluster term is equivalent to the approximation $ \langle \prod_\alpha (1+ f_\alpha ) \rangle_\perp \approx \prod_\alpha (1 + \langle f_\alpha \rangle_\perp)$.
For hard spheres $\langle f_{\alpha} \rangle_{\perp}$  can be evaluated explicitly
\begin{equation}
 {\cal V}_\text{eff}^{(2)}(r_{ij};L) = -  2 k_B T  \ln( 1 - \sqrt{ (\sigma^2 - r_{ij}^2)/L^2 } ),
\end{equation}
for $\sigma_{L}\leq r_{ij}\leq \sigma$ and zero otherwise. The total pair potential ${\cal V}_{\text{total}}(r)={\cal W}(r) + {\cal V}_\text{eff}^{(2)}(r)$,  represented in Fig.~\ref{fig:soft_potential},  smoothly interpolates between the hard-core repulsion of disks with diameter $\sigma_{L}$ and the force-free region for  $r> \sigma$. The additional effective interaction
diverges logarithmically for $r\downarrow \sigma_{L}$  and approaches  zero as a  square root  $\sqrt{(\sigma^2 -r^2)/L^2}$ for $r\uparrow \sigma$.

The effective potential can be used for the calculation of ensemble  averages. Consider an observable $X=X(\{ \vec{r}_i \})$ which depends only on the lateral coordinates. Then its configurational average yields
\begin{align}
 \frac{1}{Z} {\sf Tr} \left[ X \exp(- \beta V) \right] =& \frac{Z_\perp Z_\parallel}{Z} \langle X \langle \prod_\alpha (1+ f_\alpha ) \rangle_\perp \rangle_\parallel \nonumber \\
=& \text{e}^{\beta \Delta F} \langle X \text{e}^{-\beta V_\text{eff} } \rangle_\parallel,
\end{align}
where $\Delta F :=F-F_{\text{id}}-F_{\text{ex}}^{\parallel} = -k_{B}T \ln \langle \exp (-\beta V_\text{eff}) \rangle_{\parallel}$ is the shift of the free energy induced by the coupling between transversal and lateral d.o.f., and $\langle \cdot \rangle_\parallel$ indicates a configurational average with respect to $\rho_\parallel$.
Hence the knowledge of the effective potential allows calculating all structural quantities of the lateral d.o.f. in the confined system. In particular, using the effective  two-body potential yields averages that are correct  including  to order ${\cal O}(n L^2)$.

The shift of  the  free energy $\Delta F$ can now be evaluated explicitly to leading order.
Since the support of $f_{ij}$ is concentrated to   a tiny shell of width $\sim L^2$, the corrections with respect to the reference system become small and induce a hierarchy of contributions of decreasing weight. Then the average $\langle \exp(-\beta V_\text{eff} ) \rangle_\parallel$, c.f.  Eq. \eqref{eq:f}, evaluates to  a power series in $L^2$.
Abbreviating $y =  \langle \langle \prod_\alpha (1+ f_\alpha) \rangle_\perp \rangle_\parallel - 1$,   yields $-\beta \Delta F = \sum_{k=1}^\infty (-1)^{k+1} y^k/k$. Since
 $y = {\cal O}(L^2)$ , we find for the leading correction of the free energy
\begin{equation}\label{eq:Delta}
 -\beta \Delta F= \sum_\alpha \langle \langle f_\alpha \rangle_\perp \rangle_\parallel  + {\cal O}(L^4).
\end{equation}
The leading correction arises from  the effective pair potential $\mathcal{V}^{(2)}_{\text{eff}}$.  In the thermodynamic limit one infers
\begin{equation}
 \Delta F/N = -\frac{n k_B T}{2} \int g(r) [ \text{e}^{- \beta {\cal V}_\text{eff}^{(2)}(r;L) } -1 ]  \diff^2 r + {\cal O}( n L^2 )^2,
\end{equation}
where $g(r)$ is the radial pair distribution function of the hard-disk reference fluid.
Since the effective potential acts only on the thin layer, $g(r)$ can be replaced by its contact value $g(\sigma_L^+)$
and the remaining integral be performed with the result
\begin{equation}
 \Delta F/N =  \frac{5}{12} \pi k_B T    n L^2 g(\sigma^+) + {\cal O}(n L^2 )^2.
\end{equation}
The preceding relation uncovers the smallness parameter $n L^2$ of the confinement problem, which is one of our principal results.  Let us emphasize, that this coincides only formally with a virial low-density expansion; the quality of our approximation does not arise due to a dilute system, rather by the strong confinement. The contact value is connected to the excess surface tension $\Sigma_\text{ex} = - (\partial F_\text{ex}/\partial A)_{T,N}$  of the two-dimensional reference system via the virial equation $\Sigma_\text{ex}/n k_B T =  \pi n \sigma_L^2 g(\sigma^+_L)/2$, similar to the three-dimensional
case~\cite{Hansen:Theory_of_Simple_Liquids}.

Let us discuss some consequences of these results. For instance, the force per area exerted on the plates  $p(T,L,n=N/A)=-A^{-1} (\partial F/\partial L)_{T,A,N}$ follows to
\begin{equation}
p =   \frac{n k_B T}{L}  \left[ 1  - \frac{5}{6}   \pi n L^2  g(\sigma^+) +    \mathcal{O}(n L^2)^2 \right],
\end{equation}
where the leading term is of purely entropic origin and arises  from the ideal gas term in the transversal direction. The excess free energy $F_\text{ex}^{\parallel}$ of the reference hard-disk system does not contribute and the coupling of the lateral d.o.f. to the transversal ones is evaluated to leading order.  Similarly, the surface tension
\begin{equation}
\frac{ \Sigma}{n k_B T }= 1 + \frac{\pi n \sigma_L^2}{2} g(\sigma^+_L) + \frac{5 \pi n L^2}{12}  g(\sigma^+) + {\cal O}( n L^2 )^2,
\end{equation}
consists of the corresponding surface tension of the reference hard-disk system and the corrections due to the coupling.
Since the effective diameter of the reference system $\sigma_L$ depends also on the plate distance, the correction due to the shift of the hard-disk surface tension is of the same order as the correction due to the coupling to the transversal d.o.f.

\begin{figure}
\includegraphics[angle=0,width=0.95\linewidth]{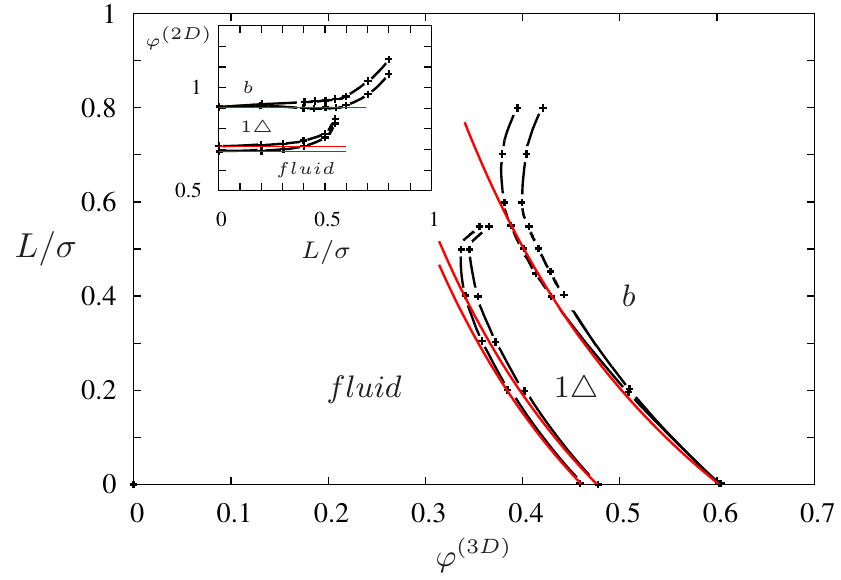}
\caption{Phase diagram of a HSHW taken from Refs.~\cite{Schmidt:1996,Schmidt:1997}. Symbols represent the MC data points, the dashed lines are guides for the eye and the thin solid lines represent~$\varphi_{*}^{(3D)}(L=0)/\varphi^{(3D)}_{*}(L)-1$ from Eq.~\eqref{eq:Phi3d}. $1\vartriangle$ and $b$ denote the triangular and buckling phase. Inset: Phase transition lines $\varphi_{*}^{(2D)}$  as function of $L/\sigma$.}
\label{fig:phase_diagram}
\end{figure}

As an application of our findings,  assume that the reference 2D fluid undergoes a phase transition at a two-dimensional packing fraction $\varphi_*^{(2D)}$. As shown above, the leading correction to the bare potential of the 2D reference  fluid is of the order $L^2$. Consequently, for small $L$  we have $\varphi_{*}^{(2D)}(L)=\varphi_{*}^{(2D)}(L=0) [1+ \mathcal{O}(L^2)] $. Note, that this remains true if besides the hard core interactions smooth pair and particle-wall interactions are added, as well as for point particles.  The result for a HSHW ~\cite{Schmidt:1996, Schmidt:1997} allows us to quantify this behavior. The 3D and 2D packing fractions are related by $\varphi^{(3D)}(L)=(2/3)\varphi^{(2D)}(L)/(1+L/\sigma)$. Then the above discussion leads to

\begin{align}
& \varphi_*^{(3D)}(L) \simeq  \varphi_*^{(3D)}(L=0)/(1+L/\sigma).
\label{eq:Phi3d}
\end{align}
Therefore, to leading order the $L$-dependence of the phase transition line $\varphi_{*}^{(3D)}(L)$  arises only from the trivial factor $1/(1+L/\sigma)$.
The corrections are ${\cal O}(L^2)$ and originate again from the coupling of transversal to lateral d.o.f.. Figure~\ref{fig:phase_diagram} shows part of the phase diagram obtained in Refs.~\cite{Schmidt:1996, Schmidt:1997} including our leading order result, Eq.~\eqref{eq:Phi3d} (see Refs. ~\cite{Xu:2008a,Xu:2008b} for a related figure). The figure demonstrates that
the freezing and melting line between fluid and triangular phase are well described by our analytic prediction  up to $ L/\sigma \lesssim 0.3$.  The freezing phase boundary between triangular and buckling phase follows  Eq.~\eqref{eq:Phi3d} even up  to $L/\sigma \lesssim 0.5$. Since the buckling phase develops a transversal structure with increasing $L/\sigma$, our cluster expansion cannot be applied, because macroscopic clusters will be involved.
The range of $L/\sigma$ for which the 2D behavior dominates becomes
even more evident using the 2D packing fraction (c.f. the inset of Fig.~\ref{fig:phase_diagram}). This inset clearly demonstrates the shallow rise of the curvature of the transition line
 with increasing $L/\sigma$ and their vanishing slope at
$L=0$, as predicted by our analytical result.  Additionally, the Monte Carlo (MC) data hint that the slope of the melting line of the b-phase is nonzero at $L=0$.

To summarize and conclude, we have proven that the lateral and transversal d.o.f. of an extremely confined fluid in slit geometry decouple if the effective width $L$ becomes much smaller than the average lateral particle distance $1/\sqrt{n}$. Since the transversal d.o.f. approach ideal gas behavior for $L\to 0$, the nontrivial thermodynamic properties are completely determined by the corresponding  2D fluid.
The leading correction to the free energy due to the residual coupling has been calculated exactly, thereby identifying $nL^2 $ as smallness parameter of the problem. The phase behavior
 in extremely small slits is close to the underlying 2D fluid and we conclude that phase transitions are robust.  Let us emphasize again, that our approach is valid for a densely packed strongly interacting system in strong contrast to an ordinary virial expansion of a dilute gas. Beyond thermodynamics all structural properties can be evaluated correctly in next-to-leading order
in an effective two-dimensional ensemble where the two-body interaction consists of a hard-disk repulsion with reduced diameter and a thin smooth repulsive layer.
Due to this mapping a simulation of the hard-sphere fluid in extreme confinement  could be replaced by a simulation of an effective 2D fluid.
Similar conclusions apply for the construction of functionals in density functional theory (see Ref.~\cite{Roth:2012}).

 Phase transition lines are analytic in the vicinity of $L=0$. Within the radius of convergence of our  cluster expansion no phase transition can occur, however  it is unclear if this radius  in general 
signals a morphological  transition. In particular, 
the analytic property  implies the existence of a critical width $L_{c}$, in case that the 2D fluid does not exhibit a single first order transition as found for a fluid of hard disks ~\cite{Weber:1995,Weber:1994} (see also the discussion in Ref.~\cite{Binder:2002}). The two transition lines  emerging from both 2D transition points have to join into a first order line at $L_{c}$, independent on whether the KTHNY scenario or that of Ref.~\cite{Bernard:2011} holds. Whether $L_{c}$ is finite, as found for the 3-state Potts model~\cite{Janke:1997} which has a continuous transition in $d=2$ and a first order transition in $d=3$ is not yet obvious. 
Recent experiments for colloidal films may support a finite critical thickness $L_{c}$~\cite{Peng:2010}.

Let us conclude by stating that it is straightforward to include a smooth wall potential and a smooth part on top of the hard-core repulsion. Then the reference ensemble still factorizes, the transversal
partition function being a product of single-particle contributions. Correspondingly the force on the plates acquires an additional contribution from the wall potential. Our framework can be applied also to point particles with smooth interaction potentials and suitable wall potentials. Then the effective potential    is proportional to the mean-square displacement $\langle (\Delta z)^2\rangle_{\perp}$ of the transversal d.o.f. which is of the order $L^2$ and the free energy shift is again of the order  $nL^2$. In particular, from our analysis one can design specific particle-wall interactions minimizing $\langle (\Delta z)^2 \rangle_{\perp}$ to stabilize a two-dimensional phase behavior,
 e.g. the hexatic phase investigated  in Ref. ~\cite{Gribova:2011}  for attractive walls.

The applicability of our approach and its results are universal for any confined $d$-dimensional fluid
where the effective width $L$ of one of the spatial extensions converges to zero. The corresponding $d$-dependent smallness parameter is given by the dimensionless quantity $nL^{d-1}$, where $n$ is now the number density of the $d-1$-dimensional fluid.


\begin{acknowledgments}
We are particularly indepted to K. Binder for several insightful and stimulating discussions. We also thank M. Schmidt for providing the original figure with the phasediagram. Discussions and correspondence with S. Dietrich, C. Holm, W. Janke, H. L{\"o}wen, P. Nielaba, M. Oettel, and M. Schmidt are gratefully acknowledged, as well.
This work has been supported by the
Deutsche Forschungsgemeinschaft DFG via the  Research Unit FOR1394 ``Nonlinear Response to
Probe Vitrification''. S. L. gratefully acknowledges the support by the Cluster of Excellence ``Engineering of Advanced Materials''
 at the University of Erlangen-Nuremberg, which is funded by the DFG
 within the framework of its ``Excellence Initiative ''.
\end{acknowledgments}


%

\end{document}